\documentclass[11pt]{article} 
\usepackage{latexsym} 
\usepackage{amssymb}
\usepackage[dvips]{graphicx}
\def \be{\begin{equation}}
\def \ee{\end{equation}}

\def\gk{\kappa}
\def\hat{\widehat} 
\newcommand{\bra}[1]{\langle #1 |}
\newcommand{\ket}[1]{| #1 \rangle}

\newcommand{\rf}[1]{(\ref{#1})}

\begin{document} 

\rightline{ITFA-2003-23} 
\rm\large 
\vskip 2cm 
\begin{center}
{\LARGE Observers and Measurements in Noncommutative Spacetimes} 
\end{center} 
\vspace{2cm}

\begin{center}
{Vyacheslav S.~Rychkov
\footnote{E-mail:  {\tt Rychkov@science.uva.nl}} \\ 
\vskip 1truecm {\it Insituut voor
   Theoretische Fysica, Universiteit van Amsterdam\\ Valckenierstraat
   65, 1018 XE Amsterdam, The Netherlands} } 
\end{center} 

\vspace{2cm}
\begin{center}
\bf Abstract
\end{center} 
\vspace{.5cm} 
We propose a "Copenhagen interpretation" for spacetime 
noncommutativity. The goal is to be able to predict results of simple 
experiments involving signal propagation directly from commutation 
relations. A model predicting an energy dependence of the speed of 
photons of the order $E/E_{Planck}$ is discussed in detail.
Such effects can be detectable by the GLAST telescope, to be 
launched in 2006.

\vspace{3cm}

PACS\quad 04.60.Pp, 02.40.Gh, 11.30.Cp, 03.65.Ta, 98.70.Rz, 11.10.Nx


\newpage

\section{Introduction}  
One of experimentally observable signatures of quantum gravity may be
a Planck-scale suppressed variation of the photon velocity
\be\label{disp}
v\approx 1+\xi E/E_{Planck}
\ee
where $\xi$ is a factor of order 1 (see~\cite{Amelino-Camelia2002} for
a review of potentially observable effects)  This effect may be detected by
observing short-duration $\gamma$-ray bursts occurring at cosmological
distances \cite{Amelino-Camelia1998}. The GLAST space telescope, to be
launched in 2006, will be sensitive enough for such an experiment
\cite{Norris1999}. In fact, already existing data on TeV $\gamma$-ray
flares in active galaxies set a bound $|\xi|\lesssim 250$ \cite{Biller1998,Schaefer}.   

In light of these exciting experimental developments, it is very
interesting to understand possible theoretical origins of
in-vacuo dispersion relations like \rf{disp}. One popular proposal is to say that quantum
gravity will result in some sort of spacetime noncommutativity,
\be
[x_\mu,x_\nu]\ne 0,
\ee
and try to relate in-vacuo dispersion to particular forms of postulated
commutation relations. Such studies usually proceed via constructing 
field theory on noncommutative spacetimes, discussing wave packets,
etc. (see e.g.~\cite{Amelino-Camelia2002-1}).

In contrast, we would like to build a theoretical scheme which would
allow to discuss signal propagation and derive dispersion relations directly
from kinematics encoded in the deformed commutation relations, without
recourse to dynamics.

First of all, this will require a careful analysis of the logical
structure of measurements in noncommutative spacetime. So far such
analysis in a form satisfactory for our purposes has not been carried out. The scheme that we discuss is very
similar in spirit to the Copenhagen interpretation of quantum
mechanics. Similarly to how the Copenhagen interpretation relies on
the existence of classical observers to interpret quantum mechanical
phenomena, we will require the existence of special-relativistic
observers who measure events happening in quantum spacetime.

Validity of the Copenhagen interpretation may be explained by strong
decoherence phenomena occurring when a microscopic quantum system
interacts with a macroscopic measuring device. 
Such decoherence will always occur when
there is a macroscopic / microscopic separation of scales. We thus believe
that it should be possible to analyze elementary experimental
consequences of quantum spacetime structure, such as Eq. \rf{disp}, in
terms of a suitable ``Copenhagen interpretation''.  

After this work has been completed and reported at a
local seminar, paper \cite{Toller} appeared, where a similar approach
to noncommutative spacetime is advocated.

\section{Observers and events}

As we have already mentioned, we assume existence of classical
inertial observers. For any two such observers we may speak of their
spacetime coordinates in each other's coordinate system, as well as of their relative velocity. These variables
will be assumed to have definite values. All inertial observers are assumed to
be equivalent. 

Our observers are classical, but spacetime will be quantum. This means
that its elementary element is not a point, but is rather described
by a state $\ket{\psi}$ in a Hilbert space $H$. We call such an
element an ``event'', $\ket{\psi}$ the event wavefunction, and $H$ the Hilbert space of events.

A possible spacetime events can be a particle decay, a particle
collision, or (the example we mostly use below) a photon emission. 
All such events will have wavefunctions associated with them.

The most elementary thing an observer can do is to observe (or
measure, we will use the two terms interchangeably) an event.
To do that, each observer is equipped with a set of quantum mechanical
Hermitean operators $\hat{x}_\mu$, one for each spacetime coordinate.
The results of the measurement of a coordinate $x_\mu$ is a real
random variable  $X$ with distribution $p(X)$ such that its moments are given
by 
\be
\int X^kp(X)\,dX =\bra{\psi}
(\hat{x}_\mu)^k \ket{\psi},\qquad k=0,1,2,\ldots
\ee
In particular, the average value and the standard deviation are given
by 
\begin{eqnarray}
\overline{X}&=&\bra{\psi} \hat{x}_\mu \ket{\psi},\\
\overline{(\Delta X)^2}&=&\bra{\psi}(\hat{x}_\mu)^2 \ket{\psi}-\bra{\psi} \hat{x}_\mu
\ket{\psi}^2.
\end{eqnarray}

Interesting things may happen when two different observers observe the
same event. To relate their viewpoints, we must have a rule which
tells us how the wavefunction transforms when the observer changes. In
other words, we must have a unitary transformation
\be
\ket{\psi_A}\stackrel{{\mathcal U}}{\longrightarrow} \ket{\psi_B}
\ee
of wavefunctions of the same event between observers A and B.

When we consider infinitesimal transformations, we see that we must
have the usual set of translation generators $\hat{P}_\mu$, rotation generators
$\hat{J}_{ij}$, and boosts $\hat{J}_{0i}$, all acting Hermiteanly on the Hilbert
space of events. Finite transformation are of course obtained by exponentiation.

It is instructive to compare the above setup with the standard Quantum
Mechanics example of quantum spin measurement, which contains
analogues of all of the above steps. Spin measurements are performed by
classical observers, each having his coordinate systems. Coordinate
systems of different observers are related by the rotation group. Each
observer measures spin components using a set of Hermitean operators  $\{\hat{S}_x,\hat{S}_y,\hat{S}_z\}$ which act on
normalized spin states $\ket{\psi}$ lying in a Hilbert space
$H$. Finally, spin states seen by different
observers are related by a unitary representation of the rotation
group. 

An important point to notice is that we may use classical relations
between different observers' coordinate systems when discussing spin
measurements. This is of course due to the fact that the discussed
effects are of order 1 (think about localizing the spin in the
$z$-direction and then measuring the $z$-component in the system rotated
by $90^\circ$) and are insensitive to possible minor uncertainties
in rotation angles. The same point of view can be applied to spacetime
noncommutativity --- we are interested in cumulative effects which have
to be insensitive to possible Planck-scale uncertainty in, say, the distance between
the two observers. 

To discuss physics, it remains to postulate commutation relations
between the measurable operators $\hat{x}_\mu$ and the change of
observer generators $\hat{P}_\mu$, $\hat{J}_{\mu\nu}$. The usual
commutative spacetime would correspond to the usual relations of
the Heisenberg and Poincar\'e algebras. In noncommutative spacetime
this structure will be deformed.
As we will see below, not all deformations are allowed on physical
grounds. 

In the rest of the paper we will mainly be concerned with
deformations of the Heisenberg algebra. The Lorentz algebra can be
included, and in the examples we will check that it is possible to
introduce Hermitean boost generators. However, at present we are unable to utilize
such inclusion in order to restrict allowed forms of noncommutativity.

\section{Example of a measurement}\label{exp}

Consider two observers situated distance $L$ from each other in the
$x_1$ direction. This means that if $\ket{\psi_A}$ is the wavefunction of
a spacetime event as seen by observer A, then the same event for
observer B will be represented by the wave function
\be\label{rel}
\ket{\psi_B}=e^{i\hat{P}_1 L}\ket{\psi_A}.
\ee
We now consider a Gedanken experiment which is supposed to model real
experiments measuring variable speed of signal propagation. Analysis
of this experiment will also be used in the next section to limit 
possible forms of deformations of the algebra.
 
Suppose that, at the origin of the coordinate system of observer A, 
a photon was emitted in direction of
observer B. The instance of emission will be represented in our
framework by a wavefunction of a spacetime ``event'' $\ket{\psi_A}$.  
By assumption,
\be\label{pos}
\bra{\psi_A}\hat{x}_\mu \ket{\psi_A}=0, \qquad \mu=0,1,2,3.
\ee

How is the energy $E$ of the photon to be represented in the wavefunction
of its emission event? We make a plausible assumption that it is to be
related to the {\it localization} of the event. More energetic photons
resolve spacetime better and must have better localized emission event
wavefunctions. Moreover, we assume that a quantitative relation holds:
\be\label{loc}
\overline{(\Delta x_\mu)^2}=\bra{\psi_A} (\hat{x}_\mu)^2 \ket{\psi_A}\sim
1/E^{2}, \qquad \mu=0,1,2,3,
\ee
so that the localization of the emission event is of the order of the
wavelength of the photon.

Now, the emitted photon will propagate and somehow interact with the
noncommutative structure of spacetime (which will of course depend
on the deformed part of the algebra, which we have not discussed so
far). Details of this process cannot be determined without discussing
dynamics of fields in noncommutative spacetimes. Eventually, the
photon arrives at the position of observer B and is detected by him. 
It is this process of detection of the emitted photon that constitutes an act of
observing its emission events by observer B.

Our main postulate is that whatever the process of propagation is, it
must be consistent with the generator action as expressed
by Eq.~\rf{rel}. In particular, this means that all observations
performed by observer B will tell him that the photon
was emitted at (average) time
\be\label{time}
\overline{t}=\bra{\psi_B} \hat{x}_0 \ket{\psi_B}.
\ee

Suppose now that {\it two} photons of {\it different} energies
$E_{1,2}$ were emitted, with wavefunctions $\ket{1_A}$, $\ket{2_A}$ of
their emission events satisfying \rf{pos}. Thus, for observer A the
emission happened at the same spacetime point. However, emission times
as perceived by observer B are given by \rf{time} and will in general
be different.   
Experimentally, this means that there will be {\it time delay}
in arrival of the photons
\be 
\overline{\Delta t}=\overline{t_1}-\overline{t_2}.
\ee
The actual size of this delay will, as we will see later, be determined by the commutator
$[\hat{x}_0,\hat{P}_1]$. Thus, we proceed to discuss deformations of the
algebra.  

\section{Allowed deformations of the algebra}

As we have already mentioned, we do not have any logical arguments
which would restrict allowed deformations of the Lorentz part of the
algebra (beyond obvious requirements of hermiticity and the Jacobi identities).

However, some restrictions do exist in the Heisenberg sector.
Working in the this sector means that we restrict attention to
the class of observers who are at rest with respect to each other. 
In particular, the Gedanken experiment of the previous section is
still possible.

First of all, the translation generators have to commute:
\be
[\hat{P}_\mu,\hat{P}_\nu]=0.
\ee
This is a consequence of our assumption that the observers are
classical, and have definite coordinates in each other's coordinate
systems. For instance, if observer A sees an event described by the wavefunction
$\ket{\psi}$, then both wavefunctions
\be
e^{i\hat{P}_1L_1}e^{i\hat{P}_2L_2}\ket{\psi},\qquad e^{i\hat{P}_2L_2}e^{i\hat{P}_1L_1}\ket{\psi} 
\ee
correspond to the same event as seen by the observer having
coordinates $(L_1,L_2,0)$ in observer A's coordinate system. Thus the
two wavefunctions have to coincide, which implies that $\hat{P}_1$ and
$\hat{P}_2$ commute.

Further, notice a special role played by $\hat{P}_0$. Namely,
this operator does not change the world
line of the observer, but merely shifts a reference point on it. The
resulting observer is still the same. If we go back to the Gedanken
experiment, the time delay between
photons registered by observer B cannot depend on the choice of the
reference point on his world line somewhere in the past. The same 
refers to all other measurements and observations he might
perform. Mathematically, this means that $\hat{P}_0$
should have standard commutation relations with $\hat{x}_\mu$. 
Thus we require as an axiom
\be \label{NC4}
[\hat{x}_\mu,\hat{P}_0]=i\delta_{\mu 0}.
\ee

\section{Two-dimensional example}\label{ex}

All essential features of the above setup can be already seen 
in 1+1 spacetime dimensions. The algebra in this case consists of 5
operators $x$, $t$, $P_x$, $P_t$, and the boost generator $B$ (in the
rest of the paper we do not put carets over operators). 
It follows from the above discussion that we must have
\be
[P_x,P_t]=0,\qquad[x,P_t]=0,\qquad [t,P_t]=i.
\ee
We have no restrictions on commutators with $B$. For having no
preferred choice of deformation, let us just leave the
Poincar\'e symmetry undeformed
\be 
[B,P_x]=iP_t,\qquad [B,P_t]=iP_x.
\ee

The remaining 5 commutation relations $[x,t]$, $[x,P_x]$, $[t,P_x]$,
$[x,B]$, $[t,B]$ must be consistent with the hermiticity of all operators
and with the Jacobi identities. We also require that they respect the
spatial reflection symmetry, which means that the commutators must
preserve their form under the simultaneous changes $x\to -x$, $P_x\to
-P_x$, $B\to -B$, $t\to t$, $P_t\to P_t$.

One sufficiently general way to construct such algebras is to start
with the standard commutators
\be
[x,t]=0,\quad [x,P_x]=i, \quad [t,P_x]=0, \quad [x,B]=it,\quad
[t,B]=ix,
\ee
and then define Hermitean operators
\begin{eqnarray}
t^{new}&=&t+F(P_x,P_t)x+xF(P_x,P_t),\\
x^{new}&=&G(P_x,P_t)x+xG(P_x,P_t).
\end{eqnarray}
Such an Ansatz produces commutation relations consistent with axiom
\rf{NC4}. Consistency with reflection symmetry requires that
 \begin{eqnarray}
F(-P_x,P_t)&=&-F(P_x,P_t),\\
G(-P_x,P_t)&=&G(P_x,P_t).
\end{eqnarray}

For the rest of this section let us concentrate on perhaps the
simplest interesting example which results from taking
\be
F=\frac{P_x} {2\gk},\qquad G=\frac 1 2,
\ee
where $\gk\sim E_{Planck}$ is the deformation scale.
This gives commutation relations
\begin{eqnarray}
[x,t]&=&ix/\gk,\label{a1}\\{} 
[x,P_x]&=&i,\\{} 
[t,P_x]&=&iP_x/\gk,\\{}
[x,B]&=&i(t-\frac 1{2\gk}(P_x x+xP_x)),\\{}
[t,B]&=&i(x+\frac1{2\gk}(P_x t+tP_x)-\frac 1\gk P_t x -
\frac1{2\gk^2} (P_x^2 x+xP_x^2)).\label{a5}
\end{eqnarray}
By construction, this algebra can be realized in the Hilbert space of
functions $\psi(p_x,p_t)$ with operators acting by
\begin{eqnarray}
P_x&=&p_x,\label{r1}\\
P_t&=&p_t,\\
B&=&i(p_x\frac\partial{\partial p_t}+p_t\frac\partial{\partial p_x}),\\
x&=&i\frac\partial{\partial p_x},\\
t&=&i\frac\partial{\partial p_t}+\frac{i}{2\gk}(p_x\frac\partial{\partial p_x}+\frac\partial{\partial p_x}p_x).\label{r5}
\end{eqnarray}

Analogous algebras can of course be given in any spacetime
dimensions. E.g. in 4d we may put
\begin{eqnarray}
t^{new}&=&t+\frac1{2\gk}\sum_{i=1}^3(P_ix_i+x_iP_i),
\end{eqnarray}
keeping the usual $x_i$. The resulting commutation relations
\begin{eqnarray}
[x_i,t]&=&ix_i/\gk,\label{b1}\\{}
[t,P_i]&=&iP_i/\gk,\qquad{\rm etc.}\label{b2}
\end{eqnarray}
are then symmetric under 3d rotations.

\section{In-vacuo dispersion}

Let us analyze the Gedanken experiment from Section \ref{exp} in terms of
algebra \rf{a1}-\rf{a5}. In the 4d case \rf{b1}-\rf{b2} will lead to
the same observable effects.

Coordinates of the same emission event as seen by observers A and B
are related by \rf{rel} as
\be
\bra{\psi_B}x\ket{\psi_B}=\bra{\psi_A}e^{-iP_xL}xe^{iP_xL}\ket{\psi_A}.\ee
Differentiating in $L$, we get
\be
\frac
d{dL}\bra{\psi_B}x\ket{\psi_B}=\bra{\psi_A}i[x,P_x]\ket{\psi_A}=-1.
\ee
Thus 
\be
\bra{\psi_B}x\ket{\psi_B}=\bra{\psi_A}x\ket{\psi_A}-L=-L
\ee
for a photon emitted at the origin of the A's coordinate system.

The same computation for observed time produces however a nontrivial relation
\be\label{t}
\bra{\psi_B}t\ket{\psi_B}=-\frac{L}{\gk}\bra{\psi_A}P_x\ket{\psi_A},
\ee
due to the non-vanishing commutator $[t,P_x]$.

Our algebra is realized in momentum representation by \rf{r1}-\rf{r5}.
In Section \ref{exp} we found it reasonable to assume that 
\be
\bra{\psi_A}x^2\ket{\psi_A}\sim 1/E^2,
\ee
where $E$ is the photon energy.
It follows that we must have 
\be\label{P_x}
\bra{\psi_A}P_x\ket{\psi_A}\sim E,
\ee
unless some cancellation occurs due to the symmetry of the
wavefunction in $P_x$. However, such symmetry seems unlikely, because
the emission wavefunction should encode the direction in which the
photon is emitted. Thus we assume that \rf{P_x} is true.
 
Eqs. \rf{t} and \rf{P_x} now show that if two photons of energies
$E_{1,2}$ are emitted, observer B will detect (average) time-delay on
arrival of the order
\be\label{delay}
t_1-t_2\sim \frac L\gk(E_1-E_2),
\ee
which means that the speed of signal propagation varies with energy
according to \rf{disp}.

This is not, however, the end of the story. It turns out that apart
from the time-delay, our model also incorporates another possible
striking signature contemplated by quantum-gravity phenomenologists \cite{Amelino-Camelia2002}, namely spreading of the burst with energy.
That is, spacetime noncommutativity
induces intrinsic uncertainty in the arrival time of the photons.
To find this uncertainty, we need to compute $\overline{(\Delta
  t)^2}$ for observer B. 

We have 
\be
\overline{(\Delta t)^2}=\bra{\psi_B} t^2 \ket{\psi_B}-\bra{\psi_B}t 
\ket{\psi_B}^2.
\ee
To find $\bra{\psi_B} t^2 \ket{\psi_B}$, we need to differentiate
\be
 \bra{\psi_A}e^{-iP_xL}t^2e^{iP_xL} \ket{\psi_A}
\ee
in $L$ {\it twice}, each time using the commutator
$[t,P_x]=iP_x/\kappa$. 
The result is
\be
\bra{\psi_B} t^2 \ket{\psi_B}=\bra{\psi_A} t^2-\frac L{\gk}(tP_x+P_xt)+\frac{L^2}{\gk^2}P_x^2 \ket{\psi_A}.
\ee
This results in an estimate of
\be\label{unc}
\overline{(\Delta t)^2}\sim1/E^2+L/\gk+L^2E^2/\gk^2
\ee
for the uncertainty of the arrival time.

Note that the analogous computation for the coordinate measurement gives
\be
\bra{\psi_B} x^2 \ket{\psi_B}=\bra{\psi_A} x^2 \ket{\psi_A}+L^2.
\ee
It follows that
\be
\overline{(\Delta x)^2}=\bra{\psi_B} x^2 \ket{\psi_B}-\bra{\psi_B}x 
\ket{\psi_B}^2\sim1/E^2
\ee
for observer B just as for observer A. Thus the assumed form of
noncommutativity does not increase the uncertainty of the space
coordinate measurement.

The $\gamma$-ray burst observations we have in mind will involve
typical values
\be
L\sim 1 Gpc,\qquad E\sim 10 MeV,
\ee
which leads to the uncertainty
\be
\Delta t\sim L E/\gk\sim 10^{-4} s,
\ee
the other two terms in \rf{unc} being negligible.
Thus we see that the uncertainty is comparable to the average
time-delay given by \rf{delay}. 


\section{Discussion and outlook}

In this paper we set up a scheme which allows to discuss observations
in noncommutative spacetimes. The scheme involves classical
observers perceiving events by accessing their wavefunctions. The
(deformed) Poincar\'e group acts on the Hilbert space of events and is used to
relate wavefunctions of an event as seen by different observers. 

The standard commutation relations between observable operators, such
as the event coordinate operators $x_\mu$, and the Poincar\'e generators may be
deformed, as an effective description of unknown physics at the Planck
scale. However, an important consequence of logical consistency is
that the commutators with $P_0$ be undeformed. Observable effects in
signal propagation, such as in-vacuo dispersion, result from deformed
commutators between $x_\mu$ and $P_i$.

The coordinate part of the algebra that we considered as an example in
Section \ref{ex} 
\be
[x_i,t]=ix_i/\gk,\qquad [x_i,x_j]=0
\ee
is well known as the $\kappa$-Minkowski spacetime. 
This spacetime was much studied recently, especially in connection
with the Doubly Special Relativity theories \cite{Amelino-Camelia2000,Amelino-Camelia2002-3}. 
The energy-momentum sector of these DSR theories is symmetric under the action of a Hopf algebra, the
so-called $\kappa$-Poincar\'e quantum algebra. The coproduct structure
allows then for a unique reconstruction of the spacetime sector, which
turns out to be the $\kappa$-Minkowski spacetime
\cite{Majid,Amelino-Camelia99,Kowalski-Glikman01,Kowalski-Glikman2002}.
  
However, taken as a whole, our treatment of spacetime noncommutativity is
rather different from previous ones. In particular, noncommutative spacetimes
discussed in the literature (starting with the classic paper 
\cite{Snyder1947}) typically do not satisfy our constraint that $[x_\mu,P_0]=i\delta_{\mu
  0}$. Thus the observational interpretation of those types of
noncommutativity is unclear to us. It is also unclear if imposing the
coproduct structure on the algebra, like in the DSR theories, can be
justified by any arguments apart from appealing for greater symmetry
(although see \cite{Amelino-Camelia2003}).

An advantage of our scheme is that it can be used to intuitively
understand expected observable effects directly from commutation
relations. However, such kinematical arguments can of course only
give order-of-magnitude estimates. Detailed understanding of the
structure of wavefunctions of events corresponding to real physical
processes, such as photon emission, can only be produced by dynamical
considerations.

One should thus try as a next step to develop (classical) field theory
consistent with deformed commutation relations of the type discussed
in this paper. Perhaps the existing literature on field theory in
noncommutative spacetimes, such as \cite{Amelino-Camelia2002-1}, can be of help in this undertaking.

\section*{Acknowledgements} 
I am grateful to Giovanni Arcioni, Jeroen van Dongen, Erik Verlinde, Jung-Tai Yee, and especially Kostas Skenderis for useful discussions and
suggestions. This work was supported by the Stichting voor Fundamenteel Onderzoek der Materie (FOM).

\end{document}